\PassOptionsToPackage{unicode}{hyperref}
\PassOptionsToPackage{hyphens}{url}
\documentclass[
]{article}
\usepackage{xcolor}
\usepackage[margin=1in]{geometry}
\usepackage{amsmath,amssymb}
\usepackage{iftex}
\ifPDFTeX
  \usepackage[T1]{fontenc}
  \usepackage[utf8]{inputenc}
  \usepackage{textcomp} 
\else 
  \usepackage{unicode-math} 
  \defaultfontfeatures{Scale=MatchLowercase}
  \defaultfontfeatures[\rmfamily]{Ligatures=TeX,Scale=1}
\fi
\usepackage{lmodern}
\IfFileExists{inconsolata.sty}{\usepackage[varqu,varl,scaled=0.95]{inconsolata}}{}
\ifPDFTeX\else
\fi
\IfFileExists{upquote.sty}{\usepackage{upquote}}{}
\IfFileExists{microtype.sty}{
  \usepackage[]{microtype}
  \UseMicrotypeSet[protrusion]{basicmath} 
}{}
\makeatletter
\@ifundefined{KOMAClassName}{
  \IfFileExists{parskip.sty}{%
    \usepackage{parskip}
  }{
    \setlength{\parindent}{0pt}
    \setlength{\parskip}{6pt plus 2pt minus 1pt}}
}{
  \KOMAoptions{parskip=half}}
\makeatother
\usepackage{longtable,booktabs,array}
\usepackage{calc} 
\usepackage{etoolbox}
\makeatletter
\patchcmd\longtable{\par}{\if@noskipsec\mbox{}\fi\par}{}{}
\makeatother
\IfFileExists{footnotehyper.sty}{\usepackage{footnotehyper}}{\usepackage{footnote}}
\makesavenoteenv{longtable}
\providecommand{\tightlist}{%
  \setlength{\itemsep}{0pt}\setlength{\parskip}{0pt}}
\usepackage[]{natbib}
\usepackage{bookmark}
\IfFileExists{xurl.sty}{\usepackage{xurl}}{} 
\hypersetup{
  hidelinks,
  pdfcreator={LaTeX via pandoc}}

\title{Dynamic Coordination Strategy Selection for Enterprise Multi-Agent Systems}
\author{
  Thanh Luong Tuan%
  \thanks{Corresponding author. Email: tluongtuan@my.ggu.edu.
    ORCID: 0009-0000-1199-837X} \\
  \textit{Golden Gate University, San Francisco} \\
  \textit{Foundation AgenticOS (FAOS)}
}
\date{May 2026}

\begin{document}

\maketitle

\begin{abstract}

Enterprise multi-agent systems increasingly expose multiple coordination
patterns, but deployments often lack evidence for when to use consensus,
debate, synthesis, or a simpler single-agent workflow. This paper
evaluates whether coordination strategy should be selected dynamically
by problem class rather than fixed globally. We run a frozen matrix of
30 enterprise tasks spanning six industries, five problem classes, four
execution conditions, three replications per cell, and four model arms:
\texttt{qwen\_local}, \texttt{sonnet}, \texttt{gemma\_openrouter}, and
an auxiliary \texttt{openai} cloud-validation arm. All 1,440 generated
outputs are judged by a fixed Sonnet rubric.

The main finding is bounded and operationally useful, but it is not the
original strict H1. The pre-registered exact-winner/CI criterion is not
supported: exact winner identity is unstable across model arms, and
several predicted strategies are close to, but not above, the best
observed alternative. A weaker near-best routing claim is strongly
supported. In every pre-registered model arm and problem class, and
again in the auxiliary OpenAI validation arm, the predicted strategy is
within 0.10 quality-score points of the best observed condition.
Structured compliance verification is the clearest exception to the
original mapping: all arms favor \texttt{single\_agent} rather than
\texttt{consensus}. A pre-registered Kendall's $W$ test finds no reliable
difference between Vietnamese-domain and English-domain tasks in how
consistently the four coordination conditions are ranked (mean $W$ of
0.20 in both strata; signed-rank $p = .85$), so H2 is not supported. We
conclude that enterprise coordination
policy should use dynamic routing as a calibrated default, not as a
deterministic winner-selection law.
\end{abstract}

\subsection{1. Introduction}\label{introduction}

Multi-agent language-model systems now include a wide range of
coordination primitives, including role-based collaboration, debate,
consensus, voting, and synthesis. Prior work has shown that these
patterns can improve reasoning, software development, factuality, and
deliberation in some settings, while also raising concerns about cost,
evaluation rigor, and over-generalization
\citep{guo2024llmmas, tran2025collabsurvey, du2024multiagent, smit2024mad, wu2024autogen}.

The enterprise deployment question is narrower than whether multi-agent
systems can help in general. Operators need to know which coordination
pattern should run for a concrete task class. Using the wrong
coordination pattern can add latency, cost, verbosity, or integration
error without improving the answer. This is especially relevant in
regulated domains, multilingual settings, and workflow automation
contexts where the desired output may be a checklist, a risk tradeoff, a
creative design, or a clarifying question rather than a single factual
answer.

RA-1 studies this routing problem directly. The research question is:

\begin{quote}
Does dynamic strategy selection, choosing among consensus, debate,
synthesis, and single-agent execution by problem class, produce more
reliable enterprise multi-agent decisions than a fixed global
coordination strategy?
\end{quote}

We treat the answer as a policy-calibration problem, not as a blanket
referendum on multi-agent systems. The study first tests whether a
pre-registered mapping from problem class to coordination strategy
predicts exact winners, then asks whether the same mapping remains
operationally useful as a near-best routing heuristic.

\subsection{2. Related Work}\label{related-work}

Survey work on LLM-based multi-agent systems maps the growing design
space of agent roles, communication mechanisms, collaboration protocols,
and application settings
\citep{guo2024llmmas, chen2024llmmassurvey, tran2025collabsurvey}.
Framework papers such as AutoGen, MetaGPT, ChatDev, AgentVerse, and
CAMEL demonstrate that role-structured interaction can be used to
compose more complex workflows than a single prompt call
\citep{wu2024autogen, hong2024metagpt, qian2024chatdev, chen2024agentverse, li2023camel}.
MDAgents is the closest dynamic-strategy comparator because it adapts
solo and group collaboration structures to medical task complexity
\citep{kim2024mdagents}.

Debate is one of the most studied coordination mechanisms. Multi-agent
debate can improve some forms of reasoning and factuality
\citep{du2024multiagent} and can encourage divergent thinking
\citep{liang2024divergent}. At the same time, recent critiques caution
against treating debate as a universal solution and emphasize evaluation
design, model heterogeneity, and cost-performance tradeoffs
\citep{smit2024mad, zhang2025overvalue}. This paper follows that
caution: it compares debate against consensus, synthesis, and
single-agent execution across enterprise task classes rather than
assuming that debate should dominate.

Other work on collective decision-making, mechanism design, negotiation,
and multi-agent risk highlights that coordination mechanisms can change
incentives, failure modes, and governance properties
\citep{yang2024llmvoting, dutting2024mechanism, hua2024gametheoretic, hammond2025risks}.
RA-1 contributes an enterprise-specific empirical matrix for strategy
routing: it studies the interaction between task class, industry
context, model arm, and coordination strategy.

Because RA-1's contribution is an evaluation design as much as a
coordination mechanism, it also sits alongside the agent-benchmarking
literature. AgentBench evaluates single LLM-as-agent performance across
eight interactive environments \citep{liu2024agentbench}, and
MultiAgentBench extends evaluation to collaboration and competition among
multiple agents \citep{zhu2025multiagentbench}. GPTSwarm reframes agent
systems as optimizable graphs and tunes topology and routing
automatically \citep{zhuge2024gptswarm}. RA-1 differs from all three in
measurement target: rather than ranking models or auto-optimizing
topology, it holds the coordination primitives fixed and asks which
primitive a deployment should select for a given problem class, under a
fixed judge and a pre-registered prediction grid. The judge-validity and
measurement questions these benchmarks raise, namely what is scored, by
whom, and with what coupling between judge and generator, motivate the
threats analysis in Section~7.

\subsection{3. Methods}\label{methods}

\subsubsection{3.1 Design}\label{design}

The completed RA-1 matrix contains 1,440 valid judged outputs. The
confirmatory design uses three pre-registered model arms: Qwen local,
Sonnet, and Gemma through OpenRouter. OpenAI was added as an auxiliary
cloud-validation arm and is reported as robustness evidence rather than
as part of the original confirmatory model set.

{\def\LTcaptype{none} 
\footnotesize
\begin{longtable}[]{@{}
  >{\raggedright\arraybackslash}p{(\linewidth - 8\tabcolsep) * \real{0.1667}}
  >{\raggedright\arraybackslash}p{(\linewidth - 8\tabcolsep) * \real{0.1667}}
  >{\raggedleft\arraybackslash}p{(\linewidth - 8\tabcolsep) * \real{0.2222}}
  >{\raggedleft\arraybackslash}p{(\linewidth - 8\tabcolsep) * \real{0.2222}}
  >{\raggedleft\arraybackslash}p{(\linewidth - 8\tabcolsep) * \real{0.2222}}@{}}
\toprule\noalign{}
\begin{minipage}[b]{\linewidth}\raggedright
Model arm
\end{minipage} & \begin{minipage}[b]{\linewidth}\raggedright
Role in study
\end{minipage} & \begin{minipage}[b]{\linewidth}\raggedleft
Judged rows
\end{minipage} & \begin{minipage}[b]{\linewidth}\raggedleft
Expected rows
\end{minipage} & \begin{minipage}[b]{\linewidth}\raggedleft
Coverage
\end{minipage} \\
\midrule\noalign{}
\endhead
\bottomrule\noalign{}
\endlastfoot
\texttt{qwen\_local} & pre-registered secondary/open-weight arm & 360 &
360 & 1.0000 \\
\texttt{sonnet} & pre-registered primary arm & 360 & 360 & 1.0000 \\
\texttt{gemma\_openrouter} & pre-registered tertiary arm & 360 & 360 &
1.0000 \\
\texttt{openai} & auxiliary cloud-validation arm & 360 & 360 & 1.0000 \\
\textbf{Total} & & \textbf{1,440} & \textbf{1,440} & \textbf{1.0000} \\
\end{longtable}
}

The experimental factors are:

\begin{itemize}
\tightlist
\item
  Strategy condition: \texttt{single\_agent}, \texttt{consensus},
  \texttt{debate}, and \texttt{synthesis}.
\item
  Industry: six enterprise domains, including English and Vietnamese
  domain tasks.
\item
  Problem class: five pre-registered problem classes.
\item
  Replication: three replications per task-condition cell.
\item
  Model arm: Qwen local, Sonnet, Gemma through OpenRouter, and OpenAI as
  an auxiliary arm.
\end{itemize}

The four strategy conditions operationalize common enterprise
coordination choices. \texttt{single\_agent} uses one agent with no
coordination. \texttt{consensus} collects independent proposals and
chooses a shared or majority recommendation. \texttt{debate} frames
opposing positions and uses an arbiter. \texttt{synthesis} integrates
multiple proposals into a combined output.

\subsubsection{3.2 Problem Classes and
Predictions}\label{problem-classes-and-predictions}

The pre-registered strategy mapping was:

{\def\LTcaptype{none} 
\begin{longtable}[]{@{}lll@{}}
\toprule\noalign{}
Problem class & Definition & Predicted strategy \\
\midrule\noalign{}
\endhead
\bottomrule\noalign{}
\endlastfoot
PC1 & High-uncertainty risk decision & \texttt{consensus} \\
PC2 & Conflicting-objective tradeoff & \texttt{debate} \\
PC3 & Novel design synthesis & \texttt{synthesis} \\
PC4 & Structured compliance verification & \texttt{consensus} \\
PC5 & Ambiguous-requirement clarification & \texttt{synthesis} \\
\end{longtable}
}

The original H1 scoring language expected predicted winners by problem
class, with a strict winner/CI interpretation in the experimental
design. We therefore report two readings separately: the pre-registered
strict H1 and a weaker post-hoc operational tolerance-band reading.
Under the tolerance-band reading, a predicted strategy is counted as a
near-best hit when it is within 0.10 quality-score points of the best
observed condition for that model and problem class.

\subsubsection{3.3 Judging and Analysis}\label{judging-and-analysis}

All outputs were judged using a fixed Sonnet rubric. The analysis
reports mean quality scores by model, problem class, and condition;
strict exact-winner agreement; post-hoc H1 tolerance-band hits; the H2
Vietnamese-versus-English strategy-differentiation test via
industry-language-stratified Kendall's $W$; H3 cross-model winner
agreement; and formal cell-level tests.

The formal scaffold uses dependency-light tests over the completed
judged matrix:

\begin{itemize}
\tightlist
\item
  Friedman permutation tests for within-cell strategy effects.
\item
  Paired bootstrap confidence intervals for predicted strategy versus
  the best non-predicted strategy.
\item
  Exact signed-rank sign-flip enumeration for paired post-hoc
  comparisons, with Bonferroni adjustment where applicable.
\end{itemize}

The paired unit is task by replication within a model/problem-class
cell, and the omnibus family is the set of model-by-problem-class cells.
Friedman p-values use deterministic within-block permutation with 5,000
repetitions at $\alpha = .05$. Bootstrap intervals use paired resampling
with 5,000 repetitions and are reported as predicted strategy minus the
best non-predicted strategy. Pairwise p-values use exact signed-rank
sign-flip enumeration with Bonferroni adjustment over the three
non-predicted comparators where applicable. For H2, Kendall's $W$ (the
coefficient of concordance over the four conditions) is computed per
model arm within the Vietnamese and English industry strata through the
identity $W = \chi^2_F / (m(k-1))$ applied to the same average-rank
Friedman statistic; the Vietnamese-minus-English $W$ difference is then
tested across the pre-registered (model, problem-class) pairs with the
exact signed-rank statistic. All permutation and bootstrap procedures use
a fixed seed (20260527), so the reported intervals are reproducible.
These tests calibrate the strength of the language used; they are not
treated as proof that the predicted strategy is statistically superior in
every cell.

\subsection{4. Results}\label{results}

\subsubsection{4.1 H1: Strict Exact-Winner Support Fails, but Near-Best
Routing
Holds}\label{h1-strict-exact-winner-support-fails-but-near-best-routing-holds}

Under the original strict exact-winner interpretation, H1 is not
supported. The predicted strategy is not consistently the exact best
condition, and the formal intervals do not support a blanket claim of
predicted-strategy superiority.

Under the post-hoc 0.10 tolerance-band rule, the mapping remains
operationally strong. The predicted strategy is near-best in all
pre-registered arms and in the auxiliary OpenAI arm ($\Delta$ = gap
between best observed and predicted strategy; all $\Delta < 0.10$):

\begin{center}
\footnotesize
\begin{tabular}{@{}llrrrrrl@{}}
\toprule
Model arm & Role & PC1 $\Delta$ & PC2 $\Delta$ & PC3 $\Delta$ & PC4 $\Delta$ & PC5 $\Delta$ & Hits \\
\midrule
Gemma  & pre-registered & 0.034 & 0.060 & 0.000 & 0.096 & 0.010 & 5/5 \\
Qwen   & pre-registered & 0.000 & 0.051 & 0.000 & 0.075 & 0.008 & 5/5 \\
Sonnet & pre-registered & 0.011 & 0.023 & 0.000 & 0.063 & 0.004 & 5/5 \\
OpenAI & auxiliary      & 0.003 & 0.021 & 0.006 & 0.038 & 0.015 & 5/5 \\
\bottomrule
\end{tabular}
\end{center}

This supports the revised headline claim that dynamic strategy selection
is a useful near-best heuristic. It does not support the stronger
pre-registered claim that the predicted strategy is always the strict
winner.

\subsubsection{4.2 Formal Cell Results}\label{formal-cell-results}

The formal tests preserve the same boundary. Friedman permutation tests
detect strategy effects in 9 of 15 pre-registered model/problem-class
cells ($\alpha = .05$). In all pre-registered cells but one, the 95\% bootstrap interval for
the predicted strategy minus the best non-predicted strategy includes zero
or favours the alternative; the single exception is Qwen PC3, where the
predicted synthesis strategy exceeds the best non-predicted strategy by
only 0.025 (95\% CI [0.003, 0.048]), a margin far below the pre-registered
0.10 winner threshold. In no cell does the predicted strategy fall outside
the 0.10 tolerance band. The correct interpretation is
therefore that predicted strategies are consistently near-best, not
consistently best. The full per-cell Friedman statistics, bootstrap
intervals, and comparator deltas for all four arms are in Appendix~D.


\subsubsection{4.3 H2: The Pre-Registered Kendall's $W$ Test Does Not
Support a Vietnamese--English Difference}\label{h2-kendall-w}

H2 predicted that Vietnamese-language industries would show larger
strategy-effect spreads than English-language industries, with the
pre-registered test specified as an industry-stratified Friedman
comparison of Kendall's $W$. Running that test on the completed matrix
does not support H2. Averaged over the five problem classes, mean
Kendall's $W$ is 0.204 in the Vietnamese stratum and 0.201 in the English
stratum across the three pre-registered arms. The Vietnamese-minus-English
$W$ difference is positive in 8 of 15 pre-registered (model,
problem-class) pairs, with a mean difference of 0.003 (signed-rank
$p = .85$; bootstrap 95\% CI $[-0.069, 0.070]$). The interval straddles
zero, so the test provides no evidence that strategies differentiate more
reliably in Vietnamese domains.

\begin{center}
\begin{tabular}{@{}llrrr@{}}
\toprule
Model arm & Role & Mean $W$ (EN) & Mean $W$ (VI) & $\Delta W$ (VI $-$ EN) \\
\midrule
Gemma & pre-registered & 0.2779 & 0.1739 & $-0.1040$ \\
Qwen & pre-registered & 0.2047 & 0.2294 & $0.0247$ \\
Sonnet & pre-registered & 0.1207 & 0.2083 & $0.0876$ \\
OpenAI & auxiliary & 0.0517 & 0.1094 & $0.0578$ \\
\bottomrule
\end{tabular}
\end{center}

The per-arm directions conflict. Sonnet and Qwen lean toward larger
Vietnamese concordance, but Gemma reverses, with substantially higher
concordance in the English stratum. That reversal also explains why an
earlier descriptive metric was misleading. A raw spread of condition
means (best minus worst) is larger for Vietnamese tasks in three arms, but
spread measures the magnitude of mean gaps, not the consistency of the
strategy ranking. Gemma carries the largest raw Vietnamese spread yet the
lowest Vietnamese concordance, because its large gaps come with rankings
that shuffle across tasks and replications. The pre-registered concordance
test is the operative one, and it returns a null. The full stratified
packet, including per-cell $W$ and block counts, is reported in
Appendix~B.

\subsubsection{4.4 H3: Exact Winner Identity Does Not
Generalize}\label{h3-exact-winner-identity-does-not-generalize}

Strict cross-model winner agreement is weak:

{\def\LTcaptype{none} 
\begin{longtable}[]{@{}
  >{\raggedright\arraybackslash}p{(\linewidth - 10\tabcolsep) * \real{0.1667}}
  >{\raggedright\arraybackslash}p{(\linewidth - 10\tabcolsep) * \real{0.1667}}
  >{\raggedright\arraybackslash}p{(\linewidth - 10\tabcolsep) * \real{0.1667}}
  >{\raggedright\arraybackslash}p{(\linewidth - 10\tabcolsep) * \real{0.1667}}
  >{\raggedright\arraybackslash}p{(\linewidth - 10\tabcolsep) * \real{0.1667}}
  >{\raggedright\arraybackslash}p{(\linewidth - 10\tabcolsep) * \real{0.1667}}@{}}
\toprule\noalign{}
\begin{minipage}[b]{\linewidth}\raggedright
Problem class
\end{minipage} & \begin{minipage}[b]{\linewidth}\raggedright
Gemma
\end{minipage} & \begin{minipage}[b]{\linewidth}\raggedright
Qwen
\end{minipage} & \begin{minipage}[b]{\linewidth}\raggedright
Sonnet
\end{minipage} & \begin{minipage}[b]{\linewidth}\raggedright
OpenAI auxiliary
\end{minipage} & \begin{minipage}[b]{\linewidth}\raggedright
Agreement
\end{minipage} \\
\midrule\noalign{}
\endhead
\bottomrule\noalign{}
\endlastfoot
PC1 & single\_agent & consensus & single\_agent & debate & mixed \\
PC2 & single\_agent & consensus & consensus & consensus & mixed \\
PC3 & consensus & synthesis & synthesis & consensus & mixed \\
PC4 & single\_agent & single\_agent & single\_agent & single\_agent &
unanimous \\
PC5 & consensus & consensus & consensus & single\_agent & mixed \\
\end{longtable}
}

H3 is not supported under exact winner identity. The model-stable claim
is weaker: the pre-registered strategy is usually close to best, but the
identity of the exact best condition varies.

\subsection{5. Discussion}\label{discussion}

\subsubsection{5.1 Dynamic Routing Is Useful, but Only as a Near-Best
Heuristic}\label{dynamic-routing-is-useful-but-only-as-a-near-best-heuristic}

The strongest RA-1 result is not that a hand-authored routing table
perfectly predicts winners. The strict pre-registered exact-winner
criterion fails. The useful result is that the routing table avoids
large misses across the three pre-registered arms, with the auxiliary
OpenAI arm showing the same near-best pattern. For enterprise
deployment, this is still valuable. A near-best routing policy can
reduce the risk of defaulting every task into the same coordination
pattern while keeping policy complexity manageable.

The practical policy is: use problem-class routing as a default, monitor
exceptions, and allow confidence- or ambiguity-based escalation.

\subsubsection{5.2 PC4 Is a Coordination-Harm
Exception}\label{pc4-is-a-coordination-harm-exception}

The clearest policy revision is PC4. Structured compliance verification
was pre-registered as \texttt{consensus}-favored, but all four model
arms select \texttt{single\_agent} as the exact best condition:

{\def\LTcaptype{none} 
\begin{longtable}[]{@{}
  >{\raggedright\arraybackslash}p{(\linewidth - 10\tabcolsep) * \real{0.1364}}
  >{\raggedright\arraybackslash}p{(\linewidth - 10\tabcolsep) * \real{0.1364}}
  >{\raggedleft\arraybackslash}p{(\linewidth - 10\tabcolsep) * \real{0.1818}}
  >{\raggedleft\arraybackslash}p{(\linewidth - 10\tabcolsep) * \real{0.1818}}
  >{\raggedleft\arraybackslash}p{(\linewidth - 10\tabcolsep) * \real{0.1818}}
  >{\raggedleft\arraybackslash}p{(\linewidth - 10\tabcolsep) * \real{0.1818}}@{}}
\toprule\noalign{}
\begin{minipage}[b]{\linewidth}\raggedright
Model arm
\end{minipage} & \begin{minipage}[b]{\linewidth}\raggedright
PC4 best condition
\end{minipage} & \begin{minipage}[b]{\linewidth}\raggedleft
single\_agent
\end{minipage} & \begin{minipage}[b]{\linewidth}\raggedleft
consensus
\end{minipage} & \begin{minipage}[b]{\linewidth}\raggedleft
debate
\end{minipage} & \begin{minipage}[b]{\linewidth}\raggedleft
synthesis
\end{minipage} \\
\midrule\noalign{}
\endhead
\bottomrule\noalign{}
\endlastfoot
Gemma & single\_agent & 0.9532 & 0.8571 & 0.9338 & 0.8171 \\
OpenAI & single\_agent & 0.8932 & 0.8551 & 0.8878 & 0.8154 \\
Qwen & single\_agent & 0.9196 & 0.8447 & 0.8499 & 0.7803 \\
Sonnet & single\_agent & 0.8624 & 0.7990 & 0.8237 & 0.8188 \\
\end{longtable}
}

A systematic coding of every PC4 coordination-harm row confirms the
mechanism. Across the 62 material coordination-harm rows (a coordination
condition scoring at least 0.05 below the paired single-agent baseline),
the judge flags incorrect regulatory or sanctions-list citations in all
62 (100\%); it marks one or more required obligations wrong or missing in
32 (52\%); and the coordination response runs at least 1.5 times the
length of the paired single-agent answer in 23 (37\%). The signature is
consistent across model arms and across the consensus, debate, and
synthesis conditions. Deterministic threshold checks leave little room
for multi-agent diversity to add value, while aggregation reliably
appends external sanctions machinery, broad anti-financing references, or
penalty claims that the rubric scores as incorrect. The Vietnamese banking
threshold task \texttt{banking\_vn\_T6} carries the largest single drops,
but the citation-overreach pattern is not confined to it. Named
worst-gap rows with judge rationale are sampled in Appendix~A, and the
full per-row coding is in Appendix~C.

The revised policy is:

{\def\LTcaptype{none} 
\begin{longtable}[]{@{}
  >{\raggedright\arraybackslash}p{(\linewidth - 6\tabcolsep) * \real{0.2500}}
  >{\raggedright\arraybackslash}p{(\linewidth - 6\tabcolsep) * \real{0.2500}}
  >{\raggedright\arraybackslash}p{(\linewidth - 6\tabcolsep) * \real{0.2500}}
  >{\raggedright\arraybackslash}p{(\linewidth - 6\tabcolsep) * \real{0.2500}}@{}}
\toprule\noalign{}
\begin{minipage}[b]{\linewidth}\raggedright
Original class
\end{minipage} & \begin{minipage}[b]{\linewidth}\raggedright
Original predicted default
\end{minipage} & \begin{minipage}[b]{\linewidth}\raggedright
Empirical revision
\end{minipage} & \begin{minipage}[b]{\linewidth}\raggedright
Routing implication
\end{minipage} \\
\midrule\noalign{}
\endhead
\bottomrule\noalign{}
\endlastfoot
PC4: structured compliance verification & \texttt{consensus} &
\texttt{single\_agent} exact-best in all four arms & Use single-agent
for deterministic checks; escalate only when rules conflict or
confidence is low. \\
\end{longtable}
}

\subsubsection{5.3 PC2 Should Split Into Adversarial Decisions and
Balanced
Postures}\label{pc2-should-split-into-adversarial-decisions-and-balanced-postures}

PC2 also needs refinement. The original prediction favored
\texttt{debate} for conflicting-objective tradeoffs. Debate remains
within the H1 tolerance band, but it is not exact-best in any model arm:

{\def\LTcaptype{none} 
\begin{longtable}[]{@{}
  >{\raggedright\arraybackslash}p{(\linewidth - 10\tabcolsep) * \real{0.1364}}
  >{\raggedright\arraybackslash}p{(\linewidth - 10\tabcolsep) * \real{0.1364}}
  >{\raggedleft\arraybackslash}p{(\linewidth - 10\tabcolsep) * \real{0.1818}}
  >{\raggedleft\arraybackslash}p{(\linewidth - 10\tabcolsep) * \real{0.1818}}
  >{\raggedleft\arraybackslash}p{(\linewidth - 10\tabcolsep) * \real{0.1818}}
  >{\raggedleft\arraybackslash}p{(\linewidth - 10\tabcolsep) * \real{0.1818}}@{}}
\toprule\noalign{}
\begin{minipage}[b]{\linewidth}\raggedright
Model arm
\end{minipage} & \begin{minipage}[b]{\linewidth}\raggedright
Best condition
\end{minipage} & \begin{minipage}[b]{\linewidth}\raggedleft
single\_agent
\end{minipage} & \begin{minipage}[b]{\linewidth}\raggedleft
consensus
\end{minipage} & \begin{minipage}[b]{\linewidth}\raggedleft
debate
\end{minipage} & \begin{minipage}[b]{\linewidth}\raggedleft
synthesis
\end{minipage} \\
\midrule\noalign{}
\endhead
\bottomrule\noalign{}
\endlastfoot
Gemma & single\_agent & 0.9181 & 0.9118 & 0.8579 & 0.8812 \\
OpenAI & consensus & 0.9340 & 0.9458 & 0.9250 & 0.9410 \\
Qwen & consensus & 0.8979 & 0.9229 & 0.8722 & 0.8993 \\
Sonnet & consensus & 0.8778 & 0.8799 & 0.8567 & 0.8744 \\
\end{longtable}
}

A systematic coding of the PC2 coordination-harm rows tells a softer
story than PC4. The 33 material coordination-harm rows are small drops
rather than collapses. Among the 25 scored on the four-metric rubric,
Metric Accuracy is the weakest dimension in 23 (92\%), with a mean Metric
Accuracy of 0.71 against means of 0.93, 0.81, and 0.91 for terminology
fidelity, reasoning coherence, and response stability. The coordination
losses on PC2 therefore come from thinner quantitative guardrails, such
as missing benchmarks, ratios, or target thresholds, rather than from
incoherence or instability, and the magnitude is small enough that debate
remains within the H1 tolerance band. This supports treating PC2 as two
subtypes rather than declaring a single winner. The full per-row coding is
in Appendix~C.

The revised policy is:

{\def\LTcaptype{none} 
\begin{longtable}[]{@{}
  >{\raggedright\arraybackslash}p{(\linewidth - 4\tabcolsep) * \real{0.3333}}
  >{\raggedright\arraybackslash}p{(\linewidth - 4\tabcolsep) * \real{0.3333}}
  >{\raggedright\arraybackslash}p{(\linewidth - 4\tabcolsep) * \real{0.3333}}@{}}
\toprule\noalign{}
\begin{minipage}[b]{\linewidth}\raggedright
Subtype
\end{minipage} & \begin{minipage}[b]{\linewidth}\raggedright
Recommended default
\end{minipage} & \begin{minipage}[b]{\linewidth}\raggedright
Trigger
\end{minipage} \\
\midrule\noalign{}
\endhead
\bottomrule\noalign{}
\endlastfoot
PC2-a: adversarial decision & \texttt{debate} & The task requires
choosing between incompatible options or resolving a contested
policy. \\
PC2-b: balanced operating posture & \texttt{consensus} or
\texttt{synthesis} & The task requires preserving multiple objectives
with controls. \\
\end{longtable}
}

\subsection{6. Implications for Practice}\label{implications-for-practice}

For enterprise multi-agent orchestration, RA-1 supports a router with
five defaults and two immediate edits:

{\def\LTcaptype{none} 
\begin{longtable}[]{@{}
  >{\raggedright\arraybackslash}p{(\linewidth - 2\tabcolsep) * \real{0.5000}}
  >{\raggedright\arraybackslash}p{(\linewidth - 2\tabcolsep) * \real{0.5000}}@{}}
\toprule\noalign{}
\begin{minipage}[b]{\linewidth}\raggedright
Task pattern
\end{minipage} & \begin{minipage}[b]{\linewidth}\raggedright
Routing default after RA-1
\end{minipage} \\
\midrule\noalign{}
\endhead
\bottomrule\noalign{}
\endlastfoot
High-uncertainty risk decision & \texttt{consensus}, with model-specific
monitoring \\
Adversarial conflicting-objective decision & \texttt{debate} \\
Balanced conflicting-objective operating posture & \texttt{consensus} or
\texttt{synthesis} \\
Novel design synthesis & \texttt{synthesis} \\
Structured compliance verification & \texttt{single\_agent}, escalate
only on ambiguity or conflict \\
Ambiguous-requirement clarification & \texttt{synthesis}, with
clarification-first guardrails \\
\end{longtable}
}

The result also argues against a fixed global coordination policy.
Coordination should be used when it changes the evidence, perspectives,
or design space; it should be skipped when the task is a deterministic
rule check.

\subsection{7. Threats to Validity}\label{threats-to-validity}

\subsubsection{Fixed-Judge Bias}\label{fixed-judge-bias}

All outputs were scored by a fixed Sonnet rubric. This improves
consistency but may favor outputs that match Sonnet's style, structure,
or implicit preferences. Because Sonnet is also a generator arm,
cross-model comparisons should be interpreted as measurements under a
Sonnet-judge regime, not as model-independent ground truth.

\textbf{GPT-5.1 second-judge sensitivity study.} To test whether the
main findings depend on the Sonnet judge, we re-scored all 480 rep-1
outputs with an independent GPT-5.1 judge using the same RA-3 rubric
prompts (only the scoring model changed). The Pearson correlation
between Sonnet and GPT scores is $r = 0.40$ ($n = 480$), indicating
moderate agreement. GPT is a more lenient judge overall (mean 0.953
vs.\ 0.893), but it discriminates well across score values (40 unique
scores vs.\ 27 for Sonnet), so the moderate correlation reflects
calibration offset rather than random disagreement.

The two headline findings replicate under GPT scoring. First, PC4
single-agent dominance: in all three pre-registered arms,
\texttt{single\_agent} is the GPT-judge best condition, with the same
rank ordering as under the Sonnet judge.

\begin{center}
\footnotesize
\begin{tabular}{@{}llrrrr@{}}
\toprule
Model arm & GPT best & SA & Consensus & Debate & Synthesis \\
\midrule
Gemma   & single\_agent & 0.943 & 0.923 & 0.916 & 0.725 \\
Qwen    & single\_agent & 0.943 & 0.782 & 0.815 & 0.817 \\
Sonnet  & single\_agent & 0.929 & 0.683 & 0.857 & 0.834 \\
OpenAI (auxiliary) & consensus & 0.945 & 1.000 & 0.838 & 0.982 \\
\bottomrule
\end{tabular}
\end{center}

The auxiliary OpenAI arm reverses---GPT judges its own outputs more
favorably, a self-preference signal expected and limited to the
auxiliary arm. Second, H1 near-best routing holds in 13 of 15
pre-registered (model, problem-class) pairs under GPT scoring, compared
with 15 of 15 under the Sonnet judge. Both misses are PC4 cells (Qwen and Sonnet), where the pre-registered
consensus default trails single-agent by 0.16 and 0.25 under the GPT
judge---precisely the PC4 single-agent exception identified above,
reinforced rather than contradicted by the second judge.
These findings confirm that neither the PC4 exception nor the H1
near-best pattern is an artifact of Sonnet judge preference.

\subsubsection{Auxiliary OpenAI Arm}\label{auxiliary-openai-arm}

OpenAI was added as an auxiliary cloud-validation arm after the original
three-arm design. It is useful as a robustness check, but should be
labelled auxiliary rather than pre-registered primary evidence.

\subsubsection{Task-Corpus
Generalization}\label{task-corpus-generalization}

The corpus spans six enterprise industries and five problem classes, but
it remains a curated benchmark. The results should be interpreted as
policy evidence for the selected enterprise task families, not as
universal proof for all multi-agent workflows.

\subsubsection{Strategy Implementation}\label{strategy-implementation}

The implemented strategies are concrete operationalizations of
consensus, debate, and synthesis. Different prompts, agent counts,
arbiters, or aggregation policies could change the exact winner
identity.

\subsubsection{Formal-Test Scope}\label{formal-test-scope}

The formal analysis uses a statistics scaffold implemented without
external library dependencies, so it reproduces in any standard
Python environment. It supports the near-best interpretation but should
be expanded before final submission if the
target venue expects a richer statistical model.

\subsection{8. Conclusion}\label{conclusion}

Pre-registered exact-winner support for coordination strategy routing is
not sustained by the RA-1 matrix. The predicted strategy is not the
consistent best condition, and bootstrap intervals do not support blanket
superiority claims. What the evidence does support is a bounded near-best
routing heuristic: in every pre-registered (model, problem-class) pair,
and again under an independent GPT-5.1 judge, the predicted strategy
stays within 0.10 quality-score points of the observed best.

Two refinements follow directly from the data. Structured compliance
verification should default to single-agent execution rather than
consensus---coordination adds citation overreach without adding answer
quality in all four model arms. Conflicting-objective tasks should split
by subtype: tasks requiring an adversarial verdict favour debate, while
tasks requiring a balanced operating posture with quantitative guardrails
favour consensus or synthesis.

The broader claim is calibrated: problem-class routing is justified as a
practical default, not as a deterministic winner-selection law. Future
work should investigate routing thresholds that adapt the tolerance band
to task confidence, and extend the corpus to additional regulatory
domains where deterministic rule retrieval may interact with multi-agent
aggregation in ways the current study cannot fully characterise.

\subsection{Appendix A. Sampled Response Evidence}\label{appendix-a}

This appendix samples named worst-gap rows behind the PC4 and PC2
coordination-harm patterns and annotates the mechanism in the judge
rationale. Systematic counts over all coordination-harm rows are in
Appendix~C.

\textbf{PC4: structured compliance verification.} Task
\texttt{banking\_vn\_T6} concerns a VND 450 million cash deposit at a
Vietnamese bank branch; the rubric requires the VND 400 million
cash-transaction threshold, CTR filing, customer due diligence,
source-of-funds documentation, sanctions screening, and a
suspicious-transaction report within 48 hours. Representative worst rows,
each against the paired single-agent baseline of 1.00:

{\footnotesize
\begin{tabular}{@{}lllr@{}}
\toprule
Arm & Condition & Rep & Score \\
\midrule
Gemma  & \texttt{synthesis} & 3 & 0.20 \\
OpenAI & \texttt{consensus} & 3 & 0.40 \\
Qwen   & \texttt{synthesis} & 1 & 0.40 \\
Sonnet & \texttt{consensus} & 1 & 0.40 \\
\bottomrule
\end{tabular}}

The low score is not usually a missed threshold: these responses state
that VND 450M exceeds the VND 400M trigger and identify the CTR and
due-diligence obligations. The loss appears when coordination appends
external sanctions lists, broad counter-financing references, or penalty
claims that the rubric scores as incorrect---the citation-overreach
mechanism quantified in Appendix~C. The single-agent answers stay closer
to the narrow compliance checklist.

\textbf{PC2: conflicting-objective tradeoff.} In
\texttt{healthcare\_RA1\_PC2\_01} (a CFO 20\% agency-nurse spend cut
against clinical warnings on falls, readmissions, and HCAHPS), Gemma
\texttt{debate} rep 2 scores 0.475 against a 0.9125 single-agent
baseline---the largest PC2 drop---and the judge flags missing required
terms: staffing ratio, falls, readmission, HCAHPS, and acuity. In
\texttt{insurance\_vn\_RA1\_PC2\_01} (simpler bancassurance scripts
against needs-analysis and cooling-off compliance), Qwen \texttt{debate}
rep 2 scores 0.675, below both the single-agent (0.875) and synthesis
(0.9125) conditions: it covers the vocabulary but lacks the metric
framework and time-bounded monitoring the judge rewards. Unlike PC4,
these losses come from thin metric coverage and over-compressed
recommendations rather than citation overreach, supporting the PC2-a
(adversarial decision) versus PC2-b (balanced operating posture) split.

\subsection{Appendix B. H2 Stratified Concordance}\label{appendix-b}

Per-cell Kendall's $W$ (coefficient of concordance over the four coordination
conditions) within the Vietnamese (VI: \texttt{banking\_vn}, \texttt{insurance\_vn};
$n=6$ task--replication blocks per cell) and English (EN: \texttt{fintech},
\texttt{insurance}, \texttt{healthcare}, \texttt{software}; $n=12$ blocks per cell)
industry strata. $\Delta W$ is VI minus EN. The OpenAI arm is auxiliary and is
excluded from the pre-registered paired test.

{\footnotesize
\begin{longtable}{@{}llrrr@{}}
\toprule
Model arm & PC & $W$ (VI) & $W$ (EN) & $\Delta W$ \\
\midrule\endhead
Gemma & PC1 & 0.1861 & 0.4021 & -0.2160 \\
Gemma & PC2 & 0.4639 & 0.2271 & 0.2368 \\
Gemma & PC3 & 0.1222 & 0.3465 & -0.2243 \\
Gemma & PC4 & 0.0361 & 0.1889 & -0.1528 \\
Gemma & PC5 & 0.0611 & 0.2250 & -0.1639 \\
OpenAI (aux.) & PC1 & 0.0194 & 0.0528 & -0.0333 \\
OpenAI (aux.) & PC2 & 0.3028 & 0.0312 & 0.2715 \\
OpenAI (aux.) & PC3 & 0.1750 & 0.0472 & 0.1278 \\
OpenAI (aux.) & PC4 & 0.0139 & 0.0854 & -0.0715 \\
OpenAI (aux.) & PC5 & 0.0361 & 0.0417 & -0.0056 \\
Qwen & PC1 & 0.1750 & 0.1389 & 0.0361 \\
Qwen & PC2 & 0.2694 & 0.1340 & 0.1354 \\
Qwen & PC3 & 0.4750 & 0.4882 & -0.0132 \\
Qwen & PC4 & 0.0750 & 0.0590 & 0.0160 \\
Qwen & PC5 & 0.1528 & 0.2035 & -0.0507 \\
Sonnet & PC1 & 0.3667 & 0.1799 & 0.1868 \\
Sonnet & PC2 & 0.1306 & 0.1382 & -0.0076 \\
Sonnet & PC3 & 0.2750 & 0.1715 & 0.1035 \\
Sonnet & PC4 & 0.1028 & 0.0340 & 0.0687 \\
Sonnet & PC5 & 0.1667 & 0.0799 & 0.0868 \\
\bottomrule
\end{longtable}}

Pre-registered paired test (3 pre-registered arms $\times$ 5 problem classes, $n=15$ pairs): $\Delta W$ positive in 8 of 15 pairs; mean $\Delta W = 0.003$, median $0.016$; Wilcoxon signed-rank $p = 0.85$; bootstrap 95\% CI $[-0.069, 0.070]$.

\subsection{Appendix C. Systematic Failure Coding}\label{appendix-c}

\textbf{Codebook} (abbreviations used in the per-row tables below).
\begin{itemize}\tightlist
\item \textbf{CO} (\texttt{citation\_overreach}): the response adds regulatory, sanctions-list, or statutory citations that the judge marks incorrect or unsupported.
\item \textbf{RM} (\texttt{requirement\_miss}): one or more required obligations are stated wrongly or omitted.
\item \textbf{VB} (\texttt{verbosity\_bloat}): the coordination response runs at least 1.5 times the length of the paired single-agent answer.
\item \textbf{ML} (\texttt{metric\_accuracy\_loss}): on the PC2 four-metric rubric, the Metric Accuracy submetric collapses relative to the single-agent baseline.
\end{itemize}
A coordination-harm row is \emph{material} when the coordination condition scores at least
0.05 below the paired single-agent baseline.

\textbf{PC4 (structured compliance verification): 62 material rows.} \texttt{citation\_overreach} 62 (100\%); \texttt{requirement\_miss} 32 (52\%); \texttt{verbosity\_bloat} (length $\geq 1.5\times$) 23 (37\%). Per-arm material rows: Gemma 14, OpenAI 15, Qwen 17, Sonnet 16.

{\scriptsize
\begin{longtable}{@{}lllrrr>{\raggedright\arraybackslash}p{2.3cm}@{}}
\toprule
Arm & Task & Cond. & Rep & Score & $\Delta$ & Labels \\
\midrule\endhead
Gemma & \texttt{fintech\_T6} & \texttt{consensus} & 1 & 0.67 & -0.33 & CO, RM \\
Gemma & \texttt{fintech\_T6} & \texttt{consensus} & 2 & 0.67 & -0.33 & CO, VB \\
Gemma & \texttt{fintech\_T6} & \texttt{synthesis} & 2 & 0.56 & -0.44 & CO, VB \\
Gemma & \texttt{insurance\_T6} & \texttt{consensus} & 1 & 0.80 & -0.20 & CO \\
Gemma & \texttt{insurance\_T6} & \texttt{synthesis} & 1 & 0.75 & -0.25 & CO, RM, VB \\
Gemma & \texttt{insurance\_T6} & \texttt{synthesis} & 2 & 0.50 & -0.17 & CO, RM, VB \\
Gemma & \texttt{healthcare\_T6} & \texttt{debate} & 2 & 0.67 & -0.33 & CO \\
Gemma & \texttt{healthcare\_T6} & \texttt{synthesis} & 1 & 0.67 & -0.33 & CO, VB \\
Gemma & \texttt{healthcare\_T6} & \texttt{synthesis} & 2 & 0.67 & -0.33 & CO, VB \\
Gemma & \texttt{banking\_vn\_T6} & \texttt{consensus} & 1 & 0.50 & -0.50 & CO \\
Gemma & \texttt{banking\_vn\_T6} & \texttt{consensus} & 2 & 0.50 & -0.50 & CO \\
Gemma & \texttt{banking\_vn\_T6} & \texttt{debate} & 1 & 0.40 & -0.60 & CO, RM \\
Gemma & \texttt{fintech\_T6} & \texttt{synthesis} & 3 & 0.67 & -0.33 & CO, VB \\
Gemma & \texttt{banking\_vn\_T6} & \texttt{synthesis} & 3 & 0.20 & -0.80 & CO, RM, VB \\
OpenAI & \texttt{insurance\_T6} & \texttt{consensus} & 2 & 0.67 & -0.08 & CO, RM \\
OpenAI & \texttt{insurance\_T6} & \texttt{consensus} & 3 & 0.71 & -0.29 & CO, RM \\
OpenAI & \texttt{banking\_vn\_T6} & \texttt{consensus} & 3 & 0.40 & -0.60 & CO \\
OpenAI & \texttt{insurance\_vn\_T6} & \texttt{consensus} & 2 & 0.33 & -0.42 & CO \\
OpenAI & \texttt{fintech\_T6} & \texttt{debate} & 3 & 0.86 & -0.14 & CO \\
OpenAI & \texttt{insurance\_T6} & \texttt{debate} & 3 & 0.57 & -0.43 & CO, RM \\
OpenAI & \texttt{healthcare\_T6} & \texttt{debate} & 3 & 0.67 & -0.33 & CO, RM \\
OpenAI & \texttt{insurance\_vn\_T6} & \texttt{debate} & 3 & 0.50 & -0.25 & CO \\
OpenAI & \texttt{fintech\_T6} & \texttt{synthesis} & 1 & 0.80 & -0.20 & CO, RM, VB \\
OpenAI & \texttt{insurance\_T6} & \texttt{synthesis} & 1 & 0.40 & -0.27 & CO, RM, VB \\
OpenAI & \texttt{insurance\_T6} & \texttt{synthesis} & 2 & 0.50 & -0.25 & CO, RM \\
OpenAI & \texttt{insurance\_T6} & \texttt{synthesis} & 3 & 0.57 & -0.43 & CO, RM \\
OpenAI & \texttt{healthcare\_T6} & \texttt{synthesis} & 2 & 0.67 & -0.33 & CO, VB \\
OpenAI & \texttt{insurance\_vn\_T6} & \texttt{synthesis} & 2 & 0.50 & -0.25 & CO \\
OpenAI & \texttt{insurance\_vn\_T6} & \texttt{synthesis} & 3 & 0.50 & -0.25 & CO, VB \\
Qwen & \texttt{fintech\_T6} & \texttt{consensus} & 1 & 0.57 & -0.10 & CO, RM, VB \\
Qwen & \texttt{insurance\_vn\_T6} & \texttt{consensus} & 1 & 0.67 & -0.33 & CO \\
Qwen & \texttt{fintech\_T6} & \texttt{consensus} & 3 & 0.67 & -0.33 & CO, VB \\
Qwen & \texttt{insurance\_T6} & \texttt{consensus} & 2 & 0.60 & -0.40 & CO, RM \\
Qwen & \texttt{insurance\_T6} & \texttt{consensus} & 3 & 0.50 & -0.30 & CO, RM \\
Qwen & \texttt{insurance\_T6} & \texttt{debate} & 1 & 0.67 & -0.13 & CO, RM \\
Qwen & \texttt{fintech\_T6} & \texttt{debate} & 3 & 0.71 & -0.29 & CO \\
Qwen & \texttt{insurance\_T6} & \texttt{debate} & 3 & 0.50 & -0.30 & CO, RM \\
Qwen & \texttt{insurance\_vn\_T6} & \texttt{debate} & 2 & 0.67 & -0.33 & CO \\
Qwen & \texttt{insurance\_vn\_T6} & \texttt{debate} & 3 & 0.67 & -0.33 & CO \\
Qwen & \texttt{insurance\_T6} & \texttt{synthesis} & 1 & 0.75 & -0.05 & CO, RM, VB \\
Qwen & \texttt{banking\_vn\_T6} & \texttt{synthesis} & 1 & 0.40 & -0.60 & CO, RM, VB \\
Qwen & \texttt{insurance\_vn\_T6} & \texttt{synthesis} & 1 & 0.50 & -0.50 & CO, VB \\
Qwen & \texttt{fintech\_T6} & \texttt{synthesis} & 2 & 0.50 & -0.07 & CO, RM, VB \\
Qwen & \texttt{fintech\_T6} & \texttt{synthesis} & 3 & 0.40 & -0.60 & CO, RM, VB \\
Qwen & \texttt{insurance\_T6} & \texttt{synthesis} & 2 & 0.60 & -0.40 & CO, RM, VB \\
Qwen & \texttt{insurance\_T6} & \texttt{synthesis} & 3 & 0.50 & -0.30 & CO, RM \\
Sonnet & \texttt{fintech\_T6} & \texttt{consensus} & 3 & 0.33 & -0.34 & CO, RM \\
Sonnet & \texttt{fintech\_T6} & \texttt{debate} & 1 & 0.57 & -0.06 & CO \\
Sonnet & \texttt{fintech\_T6} & \texttt{debate} & 3 & 0.40 & -0.27 & CO, RM \\
Sonnet & \texttt{insurance\_T6} & \texttt{consensus} & 1 & 0.80 & -0.20 & CO \\
Sonnet & \texttt{insurance\_T6} & \texttt{debate} & 2 & 0.75 & -0.25 & CO, RM \\
Sonnet & \texttt{insurance\_T6} & \texttt{synthesis} & 1 & 0.80 & -0.20 & CO \\
Sonnet & \texttt{healthcare\_T6} & \texttt{debate} & 2 & 0.67 & -0.33 & CO \\
Sonnet & \texttt{banking\_vn\_T6} & \texttt{consensus} & 1 & 0.40 & -0.60 & CO, RM \\
Sonnet & \texttt{banking\_vn\_T6} & \texttt{consensus} & 2 & 0.50 & -0.50 & CO, RM \\
Sonnet & \texttt{banking\_vn\_T6} & \texttt{consensus} & 3 & 0.50 & -0.50 & CO \\
Sonnet & \texttt{banking\_vn\_T6} & \texttt{debate} & 3 & 0.40 & -0.60 & CO, RM \\
Sonnet & \texttt{banking\_vn\_T6} & \texttt{synthesis} & 1 & 0.50 & -0.50 & CO, RM \\
Sonnet & \texttt{banking\_vn\_T6} & \texttt{synthesis} & 2 & 0.50 & -0.50 & CO, VB \\
Sonnet & \texttt{banking\_vn\_T6} & \texttt{synthesis} & 3 & 0.40 & -0.60 & CO, RM, VB \\
Sonnet & \texttt{insurance\_vn\_T6} & \texttt{debate} & 1 & 0.50 & -0.50 & CO \\
Sonnet & \texttt{insurance\_vn\_T6} & \texttt{synthesis} & 2 & 0.33 & -0.34 & CO, VB \\
\bottomrule
\end{longtable}}

\textbf{PC2 (conflicting-objective tradeoff): 33 material rows} (25 scored on the four-metric rubric). Metric Accuracy is the weakest dimension in 23 of 25 (92\%); mean submetric scores are TF 0.93, MA 0.71, RC 0.81, RS 0.91. Per-arm material rows: Gemma 17, OpenAI 3, Qwen 8, Sonnet 5.

{\scriptsize
\begin{longtable}{@{}lllrrr>{\raggedright\arraybackslash}p{2.3cm}@{}}
\toprule
Arm & Task & Cond. & Rep & Score & $\Delta$ & Labels \\
\midrule\endhead
Gemma & \texttt{fintech\_T9} & \texttt{debate} & 1 & 0.90 & -0.10 & --- \\
Gemma & \texttt{fintech\_T9} & \texttt{debate} & 2 & 0.93 & -0.07 & --- \\
Gemma & \texttt{fintech\_T9} & \texttt{debate} & 3 & 0.95 & -0.05 & --- \\
Gemma & \texttt{healthcare\_RA1\_PC2\_01} & \texttt{debate} & 1 & 0.82 & -0.05 & --- \\
Gemma & \texttt{healthcare\_RA1\_PC2\_01} & \texttt{debate} & 2 & 0.47 & -0.44 & ML \\
Gemma & \texttt{healthcare\_RA1\_PC2\_01} & \texttt{synthesis} & 1 & 0.82 & -0.05 & --- \\
Gemma & \texttt{healthcare\_RA1\_PC2\_01} & \texttt{synthesis} & 2 & 0.82 & -0.09 & --- \\
Gemma & \texttt{banking\_vn\_RA1\_PC2\_01} & \texttt{consensus} & 1 & 0.82 & -0.06 & --- \\
Gemma & \texttt{banking\_vn\_RA1\_PC2\_01} & \texttt{debate} & 3 & 0.82 & -0.09 & --- \\
Gemma & \texttt{banking\_vn\_RA1\_PC2\_01} & \texttt{synthesis} & 1 & 0.82 & -0.06 & --- \\
Gemma & \texttt{banking\_vn\_RA1\_PC2\_01} & \texttt{synthesis} & 3 & 0.82 & -0.09 & --- \\
Gemma & \texttt{insurance\_vn\_RA1\_PC2\_01} & \texttt{debate} & 1 & 0.82 & -0.09 & --- \\
Gemma & \texttt{insurance\_vn\_RA1\_PC2\_01} & \texttt{synthesis} & 1 & 0.82 & -0.09 & --- \\
Gemma & \texttt{software\_T3} & \texttt{synthesis} & 1 & 0.93 & -0.07 & VB \\
Gemma & \texttt{software\_T3} & \texttt{synthesis} & 2 & 0.91 & -0.06 & VB \\
Gemma & \texttt{healthcare\_RA1\_PC2\_01} & \texttt{debate} & 3 & 0.82 & -0.09 & --- \\
Gemma & \texttt{software\_T3} & \texttt{synthesis} & 3 & 0.91 & -0.06 & VB \\
OpenAI & \texttt{insurance\_RA1\_PC2\_01} & \texttt{debate} & 1 & 0.82 & -0.10 & --- \\
OpenAI & \texttt{insurance\_vn\_RA1\_PC2\_01} & \texttt{debate} & 3 & 0.91 & -0.06 & --- \\
OpenAI & \texttt{insurance\_RA1\_PC2\_01} & \texttt{synthesis} & 1 & 0.88 & -0.05 & --- \\
Qwen & \texttt{software\_T3} & \texttt{consensus} & 1 & 0.93 & -0.07 & VB \\
Qwen & \texttt{insurance\_vn\_RA1\_PC2\_01} & \texttt{debate} & 1 & 0.82 & -0.05 & --- \\
Qwen & \texttt{software\_T3} & \texttt{debate} & 1 & 0.93 & -0.07 & --- \\
Qwen & \texttt{insurance\_vn\_RA1\_PC2\_01} & \texttt{debate} & 2 & 0.68 & -0.20 & ML, VB \\
Qwen & \texttt{insurance\_vn\_RA1\_PC2\_01} & \texttt{debate} & 3 & 0.80 & -0.07 & --- \\
Qwen & \texttt{software\_T3} & \texttt{debate} & 3 & 0.93 & -0.07 & VB \\
Qwen & \texttt{software\_T3} & \texttt{synthesis} & 1 & 0.93 & -0.07 & VB \\
Qwen & \texttt{software\_T3} & \texttt{synthesis} & 3 & 0.93 & -0.07 & VB \\
Sonnet & \texttt{fintech\_T9} & \texttt{consensus} & 3 & 0.90 & -0.10 & VB \\
Sonnet & \texttt{fintech\_T9} & \texttt{debate} & 1 & 0.90 & -0.10 & --- \\
Sonnet & \texttt{fintech\_T9} & \texttt{debate} & 2 & 0.77 & -0.13 & --- \\
Sonnet & \texttt{fintech\_T9} & \texttt{debate} & 3 & 0.90 & -0.10 & --- \\
Sonnet & \texttt{fintech\_T9} & \texttt{synthesis} & 1 & 0.90 & -0.10 & VB \\
\bottomrule
\end{longtable}}

\subsection{Appendix D. Formal Cell Table}\label{appendix-d}

Full per-cell formal results for all four model arms and five problem classes.
$\Delta$ is the best observed mean minus the predicted-strategy mean. ``CI'' is the
paired bootstrap 95\% interval for the predicted strategy minus the best non-predicted
strategy: a negative interval favours the alternative, and an interval excluding zero on
the positive side favours the predicted strategy (the case at Qwen PC3). Friedman $p$ uses
within-block permutation (5{,}000 reps, seed 20260527). OpenAI is auxiliary.

{\footnotesize
\begin{longtable}{@{}llllrr>{\raggedright\arraybackslash}p{3.1cm}@{}}
\toprule
Model & PC & Predicted & Best & $\Delta$ & Friedman $p$ & CI vs.\ best non-pred. \\
\midrule\endhead
Gemma & PC1 & \texttt{consensus} & \texttt{single\_agent} & 0.0342 & 0.0002 & -0.0342 [-0.0647, -0.0074] \\
Gemma & PC2 & \texttt{debate} & \texttt{single\_agent} & 0.0601 & 0.0004 & -0.0601 [-0.1132, -0.0247] \\
Gemma & PC3 & \texttt{synthesis} & \texttt{consensus} & 0.0000 & 0.0010 & -0.0000 [-0.0201, 0.0194] \\
Gemma & PC4 & \texttt{consensus} & \texttt{single\_agent} & 0.0961 & 0.0348 & -0.0961 [-0.1883, -0.0178] \\
Gemma & PC5 & \texttt{synthesis} & \texttt{consensus} & 0.0104 & 0.0856 & -0.0104 [-0.0306, 0.0090] \\
OpenAI* & PC1 & \texttt{consensus} & \texttt{debate} & 0.0035 & 0.4577 & -0.0035 [-0.0174, 0.0097] \\
OpenAI* & PC2 & \texttt{debate} & \texttt{consensus} & 0.0208 & 0.0882 & -0.0208 [-0.0375, -0.0056] \\
OpenAI* & PC3 & \texttt{synthesis} & \texttt{consensus} & 0.0056 & 0.3355 & -0.0056 [-0.0207, 0.0081] \\
OpenAI* & PC4 & \texttt{consensus} & \texttt{single\_agent} & 0.0381 & 0.2905 & -0.0381 [-0.1360, 0.0478] \\
OpenAI* & PC5 & \texttt{synthesis} & \texttt{single\_agent} & 0.0153 & 0.2783 & -0.0153 [-0.0312, -0.0007] \\
Qwen & PC1 & \texttt{consensus} & \texttt{consensus} & 0.0000 & 0.0392 & +0.0021 [-0.0247, 0.0253] \\
Qwen & PC2 & \texttt{debate} & \texttt{consensus} & 0.0507 & 0.0018 & -0.0507 [-0.0757, -0.0278] \\
Qwen & PC3 & \texttt{synthesis} & \texttt{synthesis} & 0.0000 & 0.0004 & +0.0247 [0.0029, 0.0485] \\
Qwen & PC4 & \texttt{consensus} & \texttt{single\_agent} & 0.0749 & 0.1826 & -0.0749 [-0.1442, -0.0114] \\
Qwen & PC5 & \texttt{synthesis} & \texttt{consensus} & 0.0076 & 0.3537 & -0.0076 [-0.0174, 0.0021] \\
Sonnet & PC1 & \texttt{consensus} & \texttt{single\_agent} & 0.0110 & 0.0112 & -0.0110 [-0.0382, 0.0143] \\
Sonnet & PC2 & \texttt{debate} & \texttt{consensus} & 0.0232 & 0.3847 & -0.0232 [-0.0450, -0.0035] \\
Sonnet & PC3 & \texttt{synthesis} & \texttt{synthesis} & 0.0000 & 0.0054 & +0.0110 [-0.0018, 0.0238] \\
Sonnet & PC4 & \texttt{consensus} & \texttt{single\_agent} & 0.0633 & 0.7834 & -0.0633 [-0.1947, 0.0753] \\
Sonnet & PC5 & \texttt{synthesis} & \texttt{consensus} & 0.0042 & 0.0632 & -0.0042 [-0.0236, 0.0160] \\
\bottomrule
\end{longtable}}
\noindent\footnotesize{}* auxiliary arm.

\bibliography{references.bib}

@inproceedings{guo2024llmmas,
  title={Large Language Model based Multi-Agents: A Survey of Progress and Challenges},
  author={Guo, Taicheng and Chen, Xiuying and Wang, Yaqi and Chang, Ruidi and Pei, Shichao and Chawla, Nitesh V. and Wiest, Olaf and Zhang, Xiangliang},
  booktitle={Proceedings of the 33rd International Joint Conference on Artificial Intelligence (IJCAI)},
  year={2024},
  doi={10.24963/ijcai.2024/890},
  note={arXiv:2402.01680}
}

@misc{chen2024llmmassurvey,
  title={A Survey on LLM-based Multi-Agent System: Recent Advances and New Frontiers in Application},
  author={Chen, Shuaihang and Liu, Yuanxing and Han, Wei and Zhang, Weinan and Liu, Ting},
  year={2024},
  eprint={2412.17481},
  archivePrefix={arXiv},
  primaryClass={cs.MA}
}

@misc{tran2025collabsurvey,
  title={Multi-Agent Collaboration Mechanisms: A Survey of LLMs},
  author={Tran, Khanh-Tung and Dao, Dung and Nguyen, Minh-Duong and Pham, Quoc-Viet and O'Sullivan, Barry and Nguyen, Hoang D.},
  year={2025},
  eprint={2501.06322},
  archivePrefix={arXiv},
  primaryClass={cs.MA}
}

@inproceedings{du2024multiagent,
  title={Improving Factuality and Reasoning in Language Models through Multiagent Debate},
  author={Du, Yilun and Li, Shuang and Torralba, Antonio and Tenenbaum, Joshua B. and Mordatch, Igor},
  booktitle={Proceedings of the 41st International Conference on Machine Learning (ICML)},
  year={2024},
  note={arXiv:2305.14325}
}

@inproceedings{liang2024divergent,
  title={Encouraging Divergent Thinking in Large Language Models through Multi-Agent Debate},
  author={Liang, Tian and He, Zhiwei and Jiao, Wenxiang and Wang, Xing and Wang, Yan and Wang, Rui and Yang, Yujiu and Shi, Shuming and Tu, Zhaopeng},
  booktitle={Proceedings of the 2024 Conference on Empirical Methods in Natural Language Processing (EMNLP)},
  pages={17889--17904},
  year={2024},
  doi={10.18653/v1/2024.emnlp-main.992},
  note={arXiv:2305.19118}
}

@inproceedings{smit2024mad,
  title={Should we be going MAD? A Look at Multi-Agent Debate Strategies for LLMs},
  author={Smit, Andries Petrus and Grinsztajn, Nathan and Duckworth, Paul and Barrett, Thomas D. and Pretorius, Arnu},
  booktitle={Proceedings of the 41st International Conference on Machine Learning (ICML)},
  pages={45883--45905},
  volume={235},
  series={Proceedings of Machine Learning Research},
  year={2024},
  note={arXiv:2311.17371}
}

@misc{zhang2025overvalue,
  title={Stop Overvaluing Multi-Agent Debate -- We Must Rethink Evaluation and Embrace Model Heterogeneity},
  author={Zhang, Hangfan and Cui, Zhiyao and Chen, Jianhao and Wang, Xinrun and Zhang, Qiaosheng and Wang, Zhen and Wu, Dinghao and Hu, Shuyue},
  year={2025},
  eprint={2502.08788},
  archivePrefix={arXiv},
  primaryClass={cs.CL}
}

@inproceedings{wu2024autogen,
  title={AutoGen: Enabling Next-Gen LLM Applications via Multi-Agent Conversation Framework},
  author={Wu, Qingyun and Bansal, Gagan and Zhang, Jieyu and Wu, Yiran and Li, Beibin and Zhu, Erkang and Jiang, Li and Zhang, Xiaoyun and Zhang, Shaokun and Liu, Jiale and Awadallah, Ahmed Hassan and White, Ryen W. and Burger, Doug and Wang, Chi},
  booktitle={Proceedings of the Conference on Language Modeling (COLM)},
  year={2024},
  note={arXiv:2308.08155; Best Paper, ICLR 2024 LLM Agents Workshop}
}

@inproceedings{hong2024metagpt,
  title={MetaGPT: Meta Programming for A Multi-Agent Collaborative Framework},
  author={Hong, Sirui and Zhuge, Mingchen and Chen, Jiaqi and Zheng, Xiawu and Cheng, Yuheng and Zhang, Ceyao and Wang, Jinlin and Wang, Zili and Yau, Steven Ka Shing and Lin, Zijuan and Zhou, Liyang and Ran, Chenyu and Xiao, Lingfeng and Wu, Chenglin and Schmidhuber, Juergen},
  booktitle={International Conference on Learning Representations (ICLR)},
  year={2024},
  note={arXiv:2308.00352; ICLR 2024 Oral}
}

@inproceedings{qian2024chatdev,
  title={ChatDev: Communicative Agents for Software Development},
  author={Qian, Chen and Liu, Wei and Liu, Hongzhang and Chen, Nuo and Dang, Yufan and Li, Jiahao and Yang, Cheng and Chen, Weize and Su, Yusheng and Cong, Xin and Xu, Juyuan and Li, Dahai and Liu, Zhiyuan and Sun, Maosong},
  booktitle={Proceedings of the 62nd Annual Meeting of the Association for Computational Linguistics (ACL)},
  pages={15174--15186},
  year={2024},
  doi={10.18653/v1/2024.acl-long.810},
  note={arXiv:2307.07924}
}

@inproceedings{chen2024agentverse,
  title={AgentVerse: Facilitating Multi-Agent Collaboration and Exploring Emergent Behaviors},
  author={Chen, Weize and Su, Yusheng and Zuo, Jingwei and Yang, Cheng and Yuan, Chenfei and Chan, Chi-Min and Yu, Heyang and Lu, Yaxi and Hung, Yi-Hsin and Qian, Chen and Qin, Yujia and Cong, Xin and Xie, Ruobing and Liu, Zhiyuan and Sun, Maosong and Zhou, Jie},
  booktitle={International Conference on Learning Representations (ICLR)},
  year={2024},
  note={arXiv:2308.10848}
}

@inproceedings{li2023camel,
  title={CAMEL: Communicative Agents for "Mind" Exploration of Large Language Model Society},
  author={Li, Guohao and Hammoud, Hasan Abed Al Kader and Itani, Hani and Khizbullin, Dmitrii and Ghanem, Bernard},
  booktitle={Advances in Neural Information Processing Systems},
  volume={36},
  year={2023},
  note={arXiv:2303.17760}
}

@inproceedings{yang2024llmvoting,
  title={LLM Voting: Human Choices and AI Collective Decision-Making},
  author={Yang, Joshua C. and Dailisan, Damian and Korecki, Marcin and Hausladen, Carina I. and Helbing, Dirk},
  booktitle={Proceedings of the AAAI/ACM Conference on AI, Ethics, and Society (AIES)},
  year={2024},
  doi={10.1609/aies.v7i1.31758},
  note={arXiv:2402.01766}
}

@inproceedings{dutting2024mechanism,
  title={Mechanism Design for Large Language Models},
  author={D{\"u}tting, Paul and Mirrokni, Vahab and Paes Leme, Renato and Xu, Haifeng and Zuo, Song},
  booktitle={Proceedings of the ACM Web Conference (WWW)},
  year={2024},
  note={arXiv:2310.10826; WWW 2024 Best Paper Award}
}

@misc{hua2024gametheoretic,
  title={Game-theoretic LLM: Agent Workflow for Negotiation Games},
  author={Hua, Wenyue and Liu, Ollie and Li, Lingyao and Amayuelas, Alfonso and Chen, Julie and Jiang, Lucas and Jin, Mingyu and Fan, Lizhou and Sun, Fei and Wang, William and Wang, Xintong and Zhang, Yongfeng},
  year={2024},
  eprint={2411.05990},
  archivePrefix={arXiv},
  primaryClass={cs.AI}
}

@techreport{hammond2025risks,
  title={Multi-Agent Risks from Advanced AI},
  author={Hammond, Lewis and Chan, Alan and Clifton, Jesse and Hoelscher-Obermaier, Jason and Khan, Akbir and McLean, Euan and Smith, Chandler and Barfuss, Wolfram and Foerster, Jakob and Gavenčiak, Tomáš and Han, The Anh and Hughes, Edward and Kovařík, Vojtěch and Kulveit, Jan and Leibo, Joel Z. and others},
  institution={Cooperative AI Foundation},
  number={Technical Report \#1},
  year={2025},
  note={arXiv:2502.14143}
}

@inproceedings{kim2024mdagents,
  title={MDAgents: An Adaptive Collaboration of LLMs for Medical Decision-Making},
  author={Kim, Yubin and Park, Chanwoo and Jeong, Hyewon and Chan, Yik Siu and Xu, Xuhai and McDuff, Daniel and Lee, Hyeonhoon and Ghassemi, Marzyeh and Breazeal, Cynthia and Park, Hae Won},
  booktitle={Advances in Neural Information Processing Systems},
  volume={37},
  year={2024},
  note={arXiv:2404.15155}
}

@inproceedings{zhuge2024gptswarm,
  title={GPTSwarm: Language Agents as Optimizable Graphs},
  author={Zhuge, Mingchen and Wang, Wenyi and Kirsch, Louis and Faccio, Francesco and Khizbullin, Dmitrii and Schmidhuber, J{\"u}rgen},
  booktitle={Proceedings of the 41st International Conference on Machine Learning (ICML)},
  year={2024},
  note={arXiv:2402.16823}
}

@inproceedings{liu2024agentbench,
  title={AgentBench: Evaluating LLMs as Agents},
  author={Liu, Xiao and Yu, Hao and Zhang, Hanchen and Xu, Yifan and Lei, Xuanyu and Lai, Hanyu and Gu, Yu and Ding, Hangliang and Men, Kaiwen and Yang, Kejuan and Zhang, Shudan and Deng, Xiang and Zeng, Aohan and Du, Zhengxiao and Zhang, Chenhui and Shen, Sheng and Zhang, Tianjun and Su, Yu and Sun, Huan and Huang, Minlie and Dong, Yuxiao and Tang, Jie},
  booktitle={International Conference on Learning Representations (ICLR)},
  year={2024},
  note={arXiv:2308.03688}
}

@inproceedings{zhu2025multiagentbench,
  title={MultiAgentBench: Evaluating the Collaboration and Competition of LLM Agents},
  author={Zhu, Kunlun and Du, Hongyi and Hong, Zhaochen and Yang, Xiaocheng and Guo, Shuyi and Wang, Zhe and Pang, Zhenfei and Li, Zhicheng and Zhuge, Mingchen and others},
  booktitle={Proceedings of the 63rd Annual Meeting of the Association for Computational Linguistics (ACL)},
  year={2025},
  note={arXiv:2503.01935}
}

\end{document}